\documentstyle[12pt]{article}
\newcommand{\beq}{\begin{equation}}
\newcommand{\eeq}{\end{equation}}

\newcommand{\beqn}{\begin{eqnarray}}
\newcommand{\eeqn}{\end{eqnarray}}
\newcommand{\dst}{&\displaystyle}
\newcommand{\n}{\mbox{${\bf n}$}}

\newcommand{\fr}[2]{\frac{#1}{#2}}
\newcommand{\p}{\mbox{${\bf p}$}}
\newcommand{\pp}{\mbox{${\bf p'}$}}
\newcommand{\s}{\mbox{${\bf S}$}}
\newcommand{\e}{\mbox{${\bf e}$}}
\newcommand{\LL}{\mbox{${\bf l}$}}
\newcommand{\en}{\mbox{${\epsilon_n}$}}
\newcommand{\al}{\mbox{${\alpha}$}}
\begin{document}

\vspace{1.0cm}
\begin{center}{\Large \bf On the squared spin-orbit correction to 
the positronium fine-structure splitting}\\
\vspace{1.0cm}      

{\bf  I.B. Khriplovich, A.I. Milstein and A.S. Yelkhovsky} \\
Budker Institute of Nuclear Physics, 630090 Novosibirsk, Russia
                         
\vspace{4.0cm}
\end{center}

\begin{abstract}                                    
In the recent paper \cite{tao} the order $\alpha^4 R_{\infty}$
corrections to the positronium $P$ levels were reconsidered. Those
calculations confirm our corresponding results, except for the
contribution due to the squared spin-orbit interaction. We present
here a new derivation of our previous result for this last
correction, this derivation being to our opinion both simple and
convincing.

\end{abstract}

\newpage

High-precision measurements of positronium level structure provide a
unique test of quantum electrodynamics. Of course, as accurate
theoretical calculations are necessary for this test. In Refs.
\cite{kmy,kmy1} we have found analytically the corrections of order
$\alpha^{4} R_{\infty}$ to the positronium $P$ levels.

As mentioned in \cite{kmy,kmy1}, this theoretical problem for states
of nonvanishing angular momenta is quite tractable, the
main difficulty being of the ``book-keeping" nature. 
 
One could be only pleased that this tedious calculation has been
checked in recent paper \cite{tao}. Our results for almost all
contributions to the correction discussed, are confirmed there. The
only exception is the term induced by the spin-orbit interaction
squared. The corresponding operator is
\beq\label{op}
V=\fr{\alpha}{8m^4}\,\fr{4\pi}{q^2}\left(\s [\p\times\pp]\right)^2.
\eeq
Here $\s$ is the total spin operator, $\p$ and $\pp$ are momenta of
the initial and final states, respectively; $q^2=(\pp-\p)^2$.
Our result for the correction induced by this operator is
\begin{equation} \label{our}
\delta E\, = \,\en \,
\fr{1}{9600}\left(1-\fr{3}{2n^2}\right)[2(\LL\s)^2+(\LL\s)+2S(S+1)].
\end{equation}
Here $\en\equiv m \al^{6}/n^{3}$, $m$ is the electron mass, $n$ is
the principal quantum number, ${\bf l}$ is the orbital angular
momentum.  The result of Ref. \cite{tao} is different:
\begin{equation} \label{tao}
\delta E\, = \,\en\,
\fr{1}{1600}\left(1-\fr{13}{12n^2}\right)[2(\LL\s)^2+(\LL\s)+2S(S+1)]
-\,\en\,\fr{1}{576}S(S+1).
\end{equation}

The disagreement is most probably due to different treatment of the
matrix element in the coordinate representation at $r\rightarrow 0$,
which is the only subtle point in the whole problem. Therefore, after
getting acquainted with Ref. \cite{tao}, we did not confine to
performing again the calculations in the coordinate representation in
two independent ways, though this check confirmed our formula
(\ref{our}) and in particular the fact that for a $P$-wave both structures
$2(\LL\s)^2+(\LL\s)$ and $2S(S+1)$ should enter with the same
coefficient. We used as well a different approach, where the most part
of calculations was done in the momentum representation, which allows
one to get rid at all of the spurious singularities at $r\rightarrow
0$, or at large $q$.  This is the approach we present below.

The momentum part of the expectation value we are interested in, 
can be written as
\beq\label{mom}
M_{km}=\fr{\alpha}{8m^4}\int\fr{d\p}{(2\pi)^3}\psi(\p)\,
\epsilon_{ikl}\,\epsilon_{jmn}p_l p_n
\,\int\fr{d\pp}{(2\pi)^3}\fr{4\pi}{q^2}\,p^{\prime}_i 
p^{\prime}_j\psi^{\prime}(\pp).
\eeq
Here
\[ \psi(\p)=\sqrt{\fr{3}{4\pi}}\,\e\p\,F(p)\;,\;\;\;
\psi^{\prime}(\pp)=\sqrt{\fr{3}{4\pi}}\,\e^{\prime}\pp\,F(p^{\prime})
\]
are the wave functions of $P$-states in the momentum representation,
${\bf e},\;\e^{\prime}$ being constant unit vectors. The integral
over ${\bf {p}^{\prime}}$ can be presented as
\beq\label{ipp}
\int\fr{d\pp}{(2\pi)^3}\fr{4\pi}{q^2}p'_i p'_j p'_k\, F(p')=
A(p)p_i p_j p_k+B(p)(\delta_{ij} p_k+\delta_{ik} p_j + \delta_{jk} p_i).
\eeq
The only term in the rhs contributing to the correction (\ref{mom}),
is $B(p)\delta_{ij} p_k$. To calculate it explicitly, let us
multiply Eq. (\ref{ipp}) by $(\delta_{ij}- p_i p_j/p^2) p_k$. In this
way we obtain
\beq\label{}
B(p)=\fr{1}{2p^2}\int\fr{d\pp}{(2\pi)^3}\fr{4\pi}{q^2}(\p\pp)
[p'\,^2-(\p\pp)^2/p^2]\,F(p').
\eeq
As to the matrix element itself, it equals now
\[ M_{km}=\fr{\alpha}{8m^4}\left(\fr{3}{4\pi}\right)\int\fr{d\p}{(2\pi)^3}
F(p)(\e\p)[\delta_{km}p^2- p_k p_m](\e^{\prime}\p)B(p) \]
\beqn\label{}
\dst
=\fr{\alpha}{80m^4}\left(\fr{1}{4\pi}\right)\int\fr{d\p}{(2\pi)^3}
\int\fr{d\pp}{(2\pi)^3}F(p)\fr{4\pi}{q^2}(\p\pp)
[p^2 p'\,^2-(\p\pp)^2]\,F(p') \nonumber\\
\dst
\times\{e_i[4\delta_{km}\delta_{ij}-\delta_{ik}\delta_{jm}
-\delta_{jk}\delta_{im}] e^{\prime}_j\}.
\eeqn
With the identity
\beq\label{}
\fr{3}{4\pi}\int d\Omega (\e \n)l_j l_i (\e^{\prime} \n)= 
\fr{3}{4\pi}\int d\Omega [\e\times\n]_i\,[\e^{\prime}\times\n]_j=
 \e\e^{\prime} \delta_{ij}- e_i e^{\prime}_j
\eeq
we arrive at the following expression for the energy correction:
\beq\label{}
\delta E=-\fr{\alpha}{320m^4}[2(\LL\s)^2+(\LL\s)+2S(S+1)]
<4\pi q^2 + \fr{4\pi}{q^2}(p^2-p'^2)^2>.
\eeq

Thus, without ever running into singular expressions or integrals, we
have demonstrated that $2(\LL\s)^2+(\LL\s)$ and $2S(S+1)$ enter the
result in a sum only. It is convenient now to go over in the
perfectly convergent radial expectation value $<4\pi q^2 +
\fr{4\pi}{q^2}(p^2-p'^2)^2>$ to the coordinate representation. With
the obvious identity $$<\p\pp>=\fr{3}{4\pi}|R'(0)|^2,$$ the
correction is rewritten as
\beq\label{}
\delta E=-\fr{\alpha}{320m^4}[2(\LL\s)^2+(\LL\s)+2S(S+1)]
\left[-6|R'(0)|^2 +\fr{m\alpha}{2}<\fr{4}{r^4}>\right].
\eeq

This is a simple matter now to reproduce our formula (\ref{our}).


\begin{thebibliography}{99}        

\bibitem{tao} Tao Zhang, Phys.Rev. A {\bf 54}, 1252 (1996).
                              
\bibitem{kmy} I.B. Khriplovich, A.I. Milstein and A.S. Yelkhovsky, Phys.
Rev.Lett. {\bf 71}, 4323 (1993).

\bibitem{kmy1} A.S. Elkhovsky, I.B. Khriplovich and A.I. Milstein,
Zh.Eksp.Teor.Fiz. {\bf 105}, 299 (1994) [Sov.Phys.JETP {\bf 78}, 159
(1994)] .




\end{thebibliography}
\end{document}